# Data-Driven Distributionally Robust Optimization for Real-Time Economic Dispatch Considering Secondary Frequency Regulation Cost

Likai Liu, *Student Member, IEEE,* Zechun Hu, *Senior Member, IEEE,* Xiaoyu Duan, *Student Member, IEEE,* and Nikhil Pathak, *Member, IEEE*

*Abstract*—With the large-scale integration of renewable power generation, frequency regulation resources (FRRs) are required to have larger capacities and faster ramp rates, which increases the cost of the frequency regulation ancillary service. Therefore, it is necessary to consider the frequency regulation cost and constraint along with real-time economic dispatch (RTED). In this paper, a data-driven distributionally robust optimization (DRO) method for RTED considering automatic generation control (AGC) is proposed. First, a Copula-based AGC signal model is developed to reflect the correlations among the AGC signal, load power and renewable generation variations. Secondly, samples of the AGC signal are taken from its conditional probability distribution under the forecasted load power and renewable generation variations. Thirdly, a distributionally robust RTED model considering the frequency regulation cost and constraint is built and transformed into a linear programming problem by leveraging the Wasserstein metric-based DRO technique. Simulation results show that the proposed method can reduce the total cost of power generation and frequency regulation.

*Index Terms*—Secondary frequency control, real-time economic dispatch, distributionally robust optimization, Copula theory.

## Nomenclature

### A. Indices & Sets
- $i$     Index of the generators, energy storage systems (ESSs), and renewable power plants.
- $k$     Index of the segments of a piecewise-linear function.
- $n$     Index of RTED time intervals.
- $b$     Index of buses.
- G     Set of all traditional generators.
- S     Set of all ESSs.
- RE     Set of all renewable power plants.

### B. Parameters
- $K$     Total number of segments of a piecewise-linear function.
- $\phi_{i,k}^{G}/\varphi_{i,k}^{G}$     The first-degree/constant term coefficient of the piecewise-linear generation cost function of generator $i$ on the $k^{th}$ segment.
- $\phi_i^{S}$     Degradation cost of ESS $i$ per MWh.
- $\phi_i^{R}$     Mileage cost of FRR $i$ per MW.
- $P_{i,n}^{RE}$     The power output of renewable power plant $i$ at the $n^{th}$ RTED time interval.
- $P_{b,n}^{LD}$     Load power of bus $b$ at the $n^{th}$ RTED time interval.
- $P_{i,\max}$     Maximum power output of generator or ESS $i$.
- $P_{i,\min}$     Minimum power output of generator or ESS $i$.
- $f_l$     Capacity of line $l$.
- $SF_{l,b}$     Shift factor of bus $b$ over line $l$.
- $\boldsymbol{B}_b^{G}$     Node-generator incidence vector of bus $b$.
- $\boldsymbol{B}_b^{S}$     Node-ESS incidence vector of bus $b$.
- $\boldsymbol{B}_b^{RE}$     Node-renewable power plant incidence vector of bus $b$.
- $\beta$     Prescribed chance constraint satisfaction probability.
- $rr_i$     Ramp rate of FRR $i$.
- $\eta_{d/c}$     Discharging/charging efficiency of the ESS (from the grid side to ESS side).
- $CE_i$     Energy capacity of ESS $i$.
- $SOC_{i,n}$     State of charge (SOC) of ESS $i$ at the end of the $n^{th}$ RTED time interval.
- $SOC_i^{\max}$     Maximum SOC of ESS $i$.
- $SOC_i^{\min}$     Minimum SOC of ESS $i$.

### C. Decision Variables
- $P_{i,n}^{G}$     Base dispatch point of generator $i$ at the $n^{th}$ RTED time interval.
- $P_{i,n}^{d/c}$     Base discharging/charging power of ESS $i$ at the $n^{th}$ RTED time interval.
- $PF_{i,n}$     Participation factor of FRR $i$ at the $n^{th}$ RTED time interval.

## I. Introduction

THE real-time power balance in current power systems is typically achieved by two coordinated processes [1]. In the first process, real-time economic dispatch (RTED) schedules generator base dispatch points according to the predictions of load power and renewable generations [2]. In the second process, the automatic generation control (AGC) system controls frequency regulation resources (FRRs) to counteract load power and renewable generation fluctuations. As the penetration of renewable power generation rises, not only do the requirements on the capacity and ramp rate of FRRs become higher but also the cost of the frequency regulation ancillary service increases [3]. Thus, RTED without considering secondary frequency regulation may be neither reliable nor economical [1].

To improve the secondary frequency regulation efficiency, researchers have proposed the idea of incorporating optimal RTED in AGC [4], [5]. Nevertheless, it is time-consuming and even impossible to solve the RTED problem in each control period of AGC. Thus, most of those studies utilized the distributed algorithm, which solves an optimization problem by distributed computing agents in an iterative way. However,

This work was supported in part by the National Key Research and Development Program of China under Grant 2016YFB0900500. *(Corresponding author: Zechun Hu.)*
The authors are with the Department of Electrical Engineering, Tsinghua University, Beijing, 100084, China (e-mail: zechhu@tsinghua.edu.cn).

merging RTED into AGC is not compatible with the current operation process adopted by the transmission system operator: RTED is executed before AGC to schedule the base dispatch point and participation factor of the FRR [6],[7]. In addition, the communication systems among computing agents in some distributed algorithms are also different from the communication systems of real transmission systems.

Instead of merging RTED into AGC, some other studies have focused on considering the secondary frequency regulation cost and corresponding technical constraint in RTED. Reddy *et al.* [8] optimized the base dispatch point and participation factor of FRR based on the forecasted net load change between two adjacent minutes. However, the modeling of minute-level power fluctuations does not match the time period of AGC, which is typically 2~8 seconds. Zhang *et al.* [1] proposed an AGC dynamics-constrained economic dispatch model that utilizes the second-level power fluctuation as a deterministic signal. However, predicting the trajectory of second-level power fluctuations several minutes into the future is difficult even impossible. This method ignores the uncertainty of AGC signals; thus, it is more suited for lookback analysis than for look-ahead dispatch. In addition, the secondary frequency regulation cost and constraint cannot be directly obtained from the power fluctuation.

To consider the frequency regulation cost and constraint in RTED, it is essential to model the uncertainty of the AGC signal. AGC signal modeling has been researched mainly for the frequency regulation service offering of the energy storage system (ESS) [9-11]. Donadee *et al.* [9] proposed a regression-based point prediction method for modeling the hourly accumulated energy of AGC signals, while Wang *et al.* [10] proposed an interval prediction method for modeling the same problem and developed a robust optimization formulation for the regulation service offering problem. Zhang *et al.* [11] utilized a $\phi$-divergence-based data-driven distributionally robust model to handle the uncertainty of AGC signals. In [9] and [10], the latest historical data were utilized in the forecasting, while in [11], all available historical data were used to estimate the mean and covariance of AGC signals, i.e., in a "one-set-fits-all" approach. However, the correlations among the AGC signal, load power and renewable generation variations were not captured in these studies. In this work, Copula theory [12] is applied to obtain the conditional probability distribution of the AGC signal under the forecasted load power and renewable generation variations.

Considering that the true distribution of AGC signals cannot be precisely forecasted, the Wasserstein metric-based distributionally robust optimization (DRO) technique is introduced in this research. The objective of DRO is to find a decision that minimizes the worst-case expected cost. The worst case is taken over an ambiguity set constructed from samples of AGC signals [13]. DRO can avoid overly conservative decisions made by robust optimization [14] and prevent the poor out-of-sample performance of stochastic programming [13]. DRO has been applied to many stochastic optimization problems, such as optimal power flow [15, 16], energy and reserve scheduling [17-19], unit commitment [20], integrated energy system operation [21], and predictive control [22]. Expecially, Polla *et al.* and Zhou *et al.* aplied the Wasserstein metric-based DRO techenique and did interesting works in the economical dispatch problem in [18] and [19]. Nevertheless, they did not consider the secondary frequency control in their economical dispatch models.

In this paper, we develop a DRO method for RTED considering secondary frequency regulation. The major contributions of this work are summarized as follows.
1) The conditional probability distribution of the AGC signal is modeled based on Copula theory, which captures the intricate but useful correlation information among the AGC signal, load power and renewable generation variations.
2) A novel data-driven framework is proposed that combines the Copula-based AGC signal model and Wasserstein metric-based DRO model. The Copula-based AGC signal model is used to generate samples for the DRO model to build the ambiguity set.
3) A distributionally robust RTED model considering the secondary frequency regulation cost and constraint is proposed and then solved by transforming it into a linear programming model using the Wasserstein metric-based DRO technique.

The remainder of the paper is organized as follows. The method for modeling AGC signals is developed in Section II. Section III builds the formulation for the RTED considering the frequency regulation cost and constraint. Section IV reformulates the RTED model as a DRO model and transforms it into a linear programming model. Numerical experiments are presented in Section V. Section VI concludes this paper.

## II. AGC SIGNAL MODELING

Secondary frequency regulation or AGC is responsible for controlling the FRRs to minimize the frequency and tie-line power deviations [23]. In every AGC period, the AGC system calculates the total regulation requirement according to the area control error and allocates the regulation command to each FRR [24].

### A. Statistical Variables of AGC Signals

AGC signals in an RTED time interval form a time series. The assessment of the frequency regulation performance is based on the overall response results of the AGC signals in one RTED time interval. For example, the PJM market requires a minimum of 75% compliance of the AGC signals [25]. Compared to an individual AGC signal in the RTED time interval, the statistical information of the AGC signals within a concerned timespan is more important and should be reflected in the AGC signal model. In addition, the AGC signal sequence is random and almost impossible to predict precisely, whereas some statistical variables derived from it are more predictable [9],[10], e.g., the accumulated energy of the AGC signal sequence. These statistical variables are defined as follows.

*1) Accumulated regulation energy*

The upward/downward regulation signal $AGC_j^{+/-}$ in the $j^{th}$ AGC period is defined as:

$$AGC_j^{+/-} \triangleq (AGC_j)^{+/-}, \quad (1)$$

where $AGC_j$ is the $j^{th}$ AGC signal, $(x)^+ \triangleq \max(x,0)$, and $(x)^- \triangleq \max(-x,0)$. The cumulative energies of upward/downward regulation signals in the $n^{th}$ RTED time interval are defined as follows:

$$E_n^{+/-} \triangleq \tau \cdot \sum_{(n-1)\cdot T < j\cdot \tau \leq n\cdot T} AGC_j^{+/-}, \quad (2)$$

where $\tau$ and $T$ denote the timespans of each AGC and RTED

*2) Accumulated regulation mileage*

The accumulated regulation mileage in the $n^{th}$ RTED time interval is calculated by:

$$M_n \triangleq \sum_{(n-1)\cdot T < j\cdot\tau \leq n\cdot T} |AGC_j - AGC_{j-1}|. \quad (3)$$

*3) Quantile of regulation amplitude*

Given the AGC signals in the $n^{th}$ RTED time interval, $MA_n^{+/-}$ denotes the quantile of regulation amplitude with proportion $\alpha_{MA}$ that the magnitudes of the upward/downward regulation signals are smaller than. The calculation of $MA_n^{+/-}$ is formulated as:

$$\sum_{(n-1)\cdot T < j\cdot\tau \leq n\cdot T} \mathcal{I}\left(AGC_j^{+/-} \leq MA_n^{+/-}\right) \cdot \frac{1}{X_n^{+/-}} = \alpha_{MA}, \quad (4)$$

where $\mathcal{I}$ is the indicator function, with $\mathcal{I}(\text{True}) = 1$ and $\mathcal{I}(\text{False}) = 0$, and $X_n^{+/-}$ is the number of upward/downward regulation signals in the AGC signal sequence.

*4) Quantile of regulation rate*

Given the AGC signal sequence in the $n^{th}$ RTED time interval, the upward/downward ramp rates of AGC signals can be calculated. Then, the quantile of regulation rate $RR_n^{+/-}$ is defined to reflect the statistical information of the ramp rates of the AGC signals. The calculation of $RR_n^{+/-}$ is similar to that of $MA_n^{+/-}$, i.e.,

$$\frac{\sum_{(n-1)\cdot T < j\cdot\tau \leq n\cdot T} \mathcal{I}\left(AGC_{j/j-1} - AGC_{j-1/j} \leq RR_n^{+/-}\right)}{\sum_{(n-1)\cdot T < j\cdot\tau \leq n\cdot T} \mathcal{I}\left(AGC_{j/j-1} - AGC_{j-1/j} > 0\right)} = \alpha_{RR}, \quad (5)$$

where the denominator is the number of upward/downward ramping AGC signals in the $n^{th}$ RTED time interval.

*Remark 1:* In the $n^{th}$ RTED time interval, the probability distribution of the AGC signal in each AGC period may be different, i.e., the distributions of $AGC_j^+$ and $AGC_{j+1}^+$ may be different. Thus, they cannot be denoted by one stochastic variable. However, the quantile of AGC signals in the $n^{th}$ RTED time interval is a single variable, which can be denoted by a single stochastic variable. This is the basic reason for using the statistical variable of the AGC signal sequence instead of an individual AGC signal to build the AGC signal model.

RTED usually adopts the rolling optimization strategy, which solves a multiperiod optimization problem at the beginning of each RTED time interval and adopts the operation strategy of the first period. We use $\xi_n^{AGC}$ to denote the vector composed of the statistical variables of AGC signals in the $n^{th}$ RTED time interval, i.e., $\xi_n^{AGC} \triangleq \{E_n^+, E_n^-, M_n, MA_n^+, MA_n^-, RR_n^+, RR_n^-\}$. We use $\xi^{AGC} = \{\xi_1^{AGC}, \ldots, \xi_N^{AGC}\}$ to denote the set of all stochastic variables in an optimization horizon, where $N$ is the number of RTED time intervals within the optimization horizon.

**B. Correlation Between AGC signals and Power Variations**

AGC signals are related to the load power and renewable generation variations, because the AGC system is mainly designed to counteract their fluctuations [23], [26]. Considering that the very-short-term forecasts of load power and renewable generation are comparatively accurate [27], the changes in their forecasted values from the start to the end of an RTED time interval are used in this study to represent their power variations. For example, the load power variation in the $n^{th}$ RTED time interval is $\Delta P_n^{LD} \triangleq \Delta P_{n+1}^{LD} - \Delta P_n^{LD}$, where $\Delta P_n^{LD}$ and $\Delta P_{n+1}^{LD}$ are the forecasted load powers at the start of the $n^{th}$ and $(n+1)^{th}$ RTED time intervals, respectively.

To illustrate the correlations among the power variations and AGC signals, historical data of a whole month from a real power system in North China are used to perform the correlation analysis. The sampling period is five minutes, i.e., every sample contains the data in the corresponding RTED time interval. A total of 8,210 samples are used for the correlation analysis after screening and removing the abnormal data points.

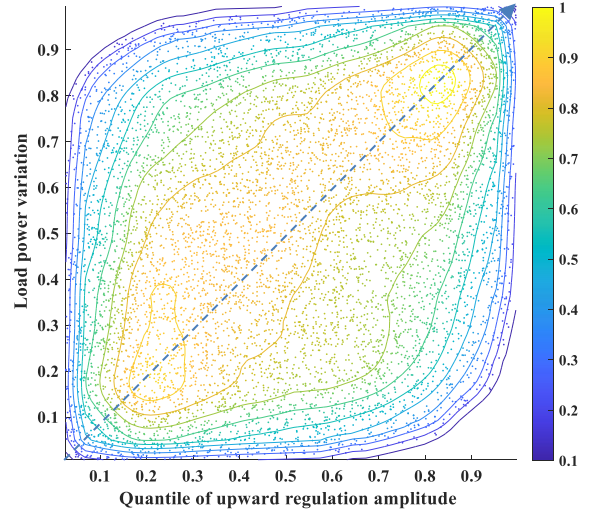

Fig. 1. Joint distributions of the load power variation and quantile of upward regulation amplitude (after CDF transformation).

Fig. 1 shows the joint distributions of the load power variation and quantile of upward regulation amplitude with $\alpha_{MA}$ equal to 70%. The original marginal distributions are transformed by the empirical cumulative distribution function (CDF) to show the correlation more clearly. The color bar in Fig. 1 indicates the normalized density of data points, and the contour line shows the zones with the same density of data points. From Fig. 1, it can be observed that the density of data points is higher in the region near the diagonal (bottom left to top right), which means that the load power variation and quantile of upward regulation amplitude are positively correlated. After the CDF transformation, a small load power variation value (Y-axis of Fig. 1) actually means a load decrease in the RTED time interval, which results in a small quantile of upward regulation amplitude. On the contrary, a large load power variation value corresponds to a load increase, contributing to a large quantile of upward regulation amplitude.

The correlation between the load power variation and the quantile of upward regulation amplitude is quantified by the Pearson correlation coefficient, which is normally used to analyze the correlation between continuous variables. The Pearson correlation coefficient between sequence $x_n \in [x_1, \ldots, x_N]$ and sequence $y_n \in [y_1, \ldots, y_N]$ is calculated by

$$\mu_{xy} = \frac{N\sum x_n y_n - \sum x_n \sum y_n}{\sqrt{N\sum x_n^2 - (\sum x_n)^2} \cdot \sqrt{N\sum y_n^2 - (\sum y_n)^2}}. \quad (6)$$

The Pearson correlation coefficient of the analyzed data is 0.32, indicating a moderate correlation between them according to [28].

**C. Copula-Based AGC Signal Modeling**

Considering the correlation between the AGC signals and power variations, their joint distribution model is built in this section. Samples of the statistical variables of AGC signals can be taken from the model according to the forecasted power variations and then used to build the ambiguity set for DRO.

Copula theory provides an effective way to construct a multivariate distribution [12], which transforms the modeling of the joint cumulative distribution into the modeling of marginal cumulative distributions and the Copula function separately. A Copula CDF and corresponding probability distribution function (PDF) are given in Appendix A as an example. By utilizing the Copula function, the joint CDF of the statistical variables of AGC signals and the variations of load power and renewable generations can be expressed as follows:

$$\mathcal{F}_{\xi_n^{\mathrm{AGC}} \Delta P_n}\left(\xi_n^{\mathrm{AGC}}, \Delta P_n\right) = \mathcal{C}_{\xi_n^{\mathrm{AGC}} \Delta P_n}\left(\mathcal{F}_{E_n^+}(E_n^+), \ldots, \mathcal{F}_{\Delta P_n^{\mathrm{WD}}}(\Delta P_n^{\mathrm{WD}})\right), \quad (7)$$

where $\Delta P_n$ is a vector composed of the load power variation $\Delta P_n^{\mathrm{LD}}$, solar power variation $\Delta P_n^{\mathrm{PV}}$, and wind power variation $\Delta P_n^{\mathrm{WD}}$ in the $n^{th}$ RTED time interval, i.e., $\Delta P_n \triangleq \{\Delta P_n^{\mathrm{LD}}, \Delta P_n^{\mathrm{PV}}, \Delta P_n^{\mathrm{WD}}\}$, $\mathcal{C}$ is the Copula function, $\mathcal{F}_{E_n^+}$ and $\mathcal{F}_{\Delta P_n^{\mathrm{WD}}}$ are the marginal CDFs of $E_n^+$ and $\Delta P_n^{\mathrm{WD}}$, respectively.

There are two key steps in modeling the joint distribution model by utilizing the Copula theory. The first step is to build the marginal CDF of each stochastic variable by fitting the historical data. Various kinds of models can be used in this step, including parametric and nonparametric distributions. The second step is to identify the key parameters of the Copula function by using the rank correlation coefficients among the stochastic variables. More details about the parameter identification of the Copula function can be found in [12]. After the CDF transformations, the arbitrarily distributed variables $E_n^+, \ldots, \Delta P_n^{\mathrm{WD}}$ become the uniformly distributed variables $\mathcal{F}_{E_n^+}(E_n^+), \ldots, \mathcal{F}_{\Delta P_n^{\mathrm{WD}}}(\Delta P_n^{\mathrm{WD}})$, and Copula theory takes advantage of the fact that stochastic dependence is more easily recognized for uniformly distributed variables in comparison to other arbitrarily distributed variables [29].

The joint PDF $f_{\xi_n^{\mathrm{AGC}} \Delta P_n}$ is the derivative of the joint CDF $\mathcal{F}_{\xi_n^{\mathrm{AGC}} \Delta P_n}$ on stochastic variables $\xi_n^{\mathrm{AGC}}$ and $\Delta P_n$, given as follows:

$$\begin{aligned} f_{\xi_n^{\mathrm{AGC}} \Delta P_n}\left(\xi_n^{\mathrm{AGC}}, \Delta P_n\right) &= \frac{\partial \mathcal{F}_{\xi_n^{\mathrm{AGC}} \Delta P_n}\left(\xi_n^{\mathrm{AGC}}, \Delta P_n\right)}{\partial \xi_n^{\mathrm{AGC}} \partial \Delta P_n} \\ &= \varsigma_{\xi_n^{\mathrm{AGC}} \Delta P_n}\left(\mathcal{F}_{E_n^+}(E_n^+), \ldots, \mathcal{F}_{\Delta P_n^{\mathrm{WD}}}(\Delta P_n^{\mathrm{WD}})\right) \\ &\quad \cdot \prod_{\xi \in \xi_n^{\mathrm{AGC}}} f_{\xi}(\xi) \cdot \prod_{\Delta P \in \Delta P_n} f_{\Delta P}(\Delta P), \end{aligned} \quad (8)$$

where $\varsigma_{\xi_n^{\mathrm{AGC}} \Delta P_n}$ is the Copula density function for the joint PDF $f_{\xi_n^{\mathrm{AGC}} \Delta P_n}$; $f_{\xi}$ and $f_{\Delta P}$ are the marginal PDFs of $\xi \in \xi_n^{\mathrm{AGC}}$ and $\Delta P \in \Delta P_n$, respectively.

Then, the conditional PDF of $\xi_n^{\mathrm{AGC}}$ under $\Delta P_n$ can be obtained by dividing the joint PDF of the power variations $f_{\Delta P_n}(\Delta P_n)$ into $f_{\xi_n^{\mathrm{AGC}} \Delta P_n}(\xi_n^{\mathrm{AGC}}, \Delta P_n)$ as

$$\begin{aligned} f_{\xi_n^{\mathrm{AGC}} | \Delta P_n}\left(\xi_n^{\mathrm{AGC}} | \Delta P_n\right) &= \frac{f_{\xi_n^{\mathrm{AGC}} \Delta P_n}\left(\xi_n^{\mathrm{AGC}}, \Delta P_n\right)}{f_{\Delta P_n}(\Delta P_n)} \\ &= \frac{\varsigma_{\xi_n^{\mathrm{AGC}} \Delta P_n}\left(\mathcal{F}_{E_n^+}(E_n^+), \ldots, \mathcal{F}_{\Delta P_n^{\mathrm{WD}}}(\Delta P_n^{\mathrm{WD}})\right)}{\varsigma_{\Delta P_n}\left(\mathcal{F}_{\Delta P_n^{\mathrm{LD}}}(\Delta P_n^{\mathrm{LD}}), \ldots, \mathcal{F}_{\Delta P_n^{\mathrm{WD}}}(\Delta P_n^{\mathrm{WD}})\right)} \cdot \prod_{\xi \in \xi_n^{\mathrm{AGC}}} f(\xi) \end{aligned} \quad (9)$$

where $f_{\Delta P_n}(\Delta P_n)$ can be calculated in a similar way as (7)-(8). Since the joint PDF of the power variations $f_{\Delta P_n}(\Delta P_n)$ is part of $f_{\xi_n^{\mathrm{AGC}} \Delta P_n}(\xi_n^{\mathrm{AGC}}, \Delta P_n)$, the marginal CDFs $\{\mathcal{F}_{\Delta P_n^{\mathrm{LD}}}, \ldots, \mathcal{F}_{\Delta P_n^{\mathrm{WD}}}\}$, marginal PDFs $\{f_{\Delta P_n^{\mathrm{LD}}}, \ldots, f_{\Delta P_n^{\mathrm{WD}}}\}$, and Copula density function $\varsigma_{\Delta P_n}$ used to calculate $f_{\Delta P_n}(\Delta P_n)$ can be obtained directly from $f_{\xi_n^{\mathrm{AGC}} \Delta P_n}(\xi_n^{\mathrm{AGC}}, \Delta P_n)$.

## III. STOCHASTIC RTED OPTIMIZATION MODEL

### A. Framework for the Data-Driven DRO Method

According to the forecasted load power and renewable generations, the power variations in the RTED time interval can be calculated. Then, samples of the statistical variables of AGC signals can be taken from conditional PDF (9) using the calculated power variations as the condition. These samples are used in the DRO model to construct the ambiguity set, and the forecasted load power as well as renewable generations are used in the RTED optimization. The Copula-based AGC signal model and Wasserstein-metric based DRO compose the data-driven DRO method, as illustrated in Fig. 2.

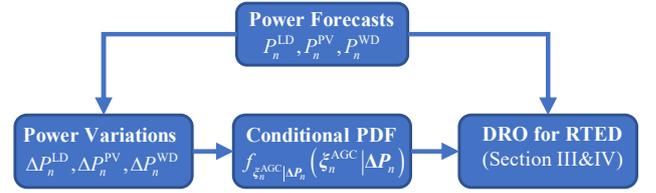

Fig. 2 Framework of the data-driven DRO method.

The rolling optimization model for RTED considering the frequency regulation cost and constraint is built in the remainder of this section.

### B. Objective Function Formulation

The objective function is composed of generation and frequency regulation costs. The regulation cost is typically composed of capacity and mileage costs [30],[31]. Only the mileage cost is considered in this study because the FRR capacity is cleared before RTED. The participation factor, which is one of the most commonly used regulation command allocation strategies [6],[7], is utilized in this research. The objective function is given as follows:

$$\min \sum_{n=1}^{N}\left\{\sum_{i \in \mathrm{S}} \Phi_{i,n}^{\mathrm{S}}\left(P_{i,n}^{\mathrm{c}}, P_{i,n}^{\mathrm{d}}\right) + \mathbb{E}^{\mathbb{P}}\left[\sum_{i \in \mathrm{G}} \Phi_{i,n}^{\mathrm{G}}\left(\overline{P_{i,n}^{\mathrm{G}}}\right) + \sum_{i \in \{\mathrm{G},\mathrm{S}\}} \Phi_{i,n}^{\mathrm{R}}\left(M_n, PF_{i,n}\right)\right]\right\} \quad (10)$$

$$\Phi_{i,n}^{\mathrm{S}}\left(P_{i,n}^{\mathrm{c}}, P_{i,n}^{\mathrm{d}}\right) = \phi_i^{\mathrm{S}} \cdot \left(P_{i,n}^{\mathrm{d}} \eta_{\mathrm{d}}^{-1} + P_{i,n}^{\mathrm{c}} \eta_{\mathrm{c}}\right) \cdot T, \quad \forall i \in \mathrm{S}, n, \quad (11)$$

$$\Phi_{i,n}^{\mathrm{G}}\left(P_{i,n}^{\mathrm{G}}\right) = \max_{k \leq K}\left(\phi_{i,k}^{\mathrm{G}} \cdot P_{i,n}^{\mathrm{G}} + \varphi_{i,k}^{\mathrm{G}}\right) \cdot T, \quad \forall i \in \mathrm{G}, n, \quad (12)$$

$$\overline{P_{i,n}^{\mathrm{G}}} = P_{i,n}^{\mathrm{G}} + \frac{PF_{i,n} \cdot \left(E_n^+ - E_n^-\right)}{T}, \quad \forall i \in \mathrm{G}, n, \quad (13)$$

$$\Phi_{i,n}^{\mathrm{R}}\left(M_n, PF_{i,n}\right) = \phi_i^{\mathrm{R}} \cdot PF_{i,n} \cdot M_n, \quad \forall i \in \{\mathrm{G},\mathrm{S}\}, n, \quad (14)$$

Equation (11) is the degradation cost $\Phi^{\mathrm{S}}$ of the ESS. The ESS devices considered in this research are battery ESSs, and we assume that the degradation cost is proportional to the charging and discharging energies, as in [10], [32]. Equation (12) represents the generation cost $\Phi^{\mathrm{G}}$ of the generator, which is calculated by the piecewise linear function. The energy consumed by the frequency regulation is considered to calculate the average output power $\overline{P^{\mathrm{G}}}$ of the generator in the RTED time interval, as in (13). Equation (14) denotes the mileage cost $\Phi^{\mathrm{R}}$ of the FRR, which is calculated according to the unit mileage cost and accumulated regulation mileage [33].

## C. System Constraints Formulation

### 1) RTED constraints

The power balance constraint is given by (15) without considering active power losses. The power range constraint for the generator and ESS is given by (16). The ramp rate constraints between adjacent RTED operation points are shown in (17). Constraint (18) is the line capacity constraint. To simplify the mathematical expression, the power output of the $i^{th}$ ESS is represented by $P_i^S$, and $P_i^S = P_i^d - P_i^c$.

$$\sum_{i \in G} P_{i,n}^G + \sum_{i \in S} P_{i,n}^S + \sum_{i \in RE} P_{i,n}^{RE} = \sum_b P_{b,n}^{LD}, \quad \forall n, \quad (15)$$

$$P_{i,\min} \leq P_{i,n}^{G/S} \leq P_{i,\max}, \quad \forall i \in \{G,S\}, n, \quad (16)$$

$$-rr_i \cdot T \leq P_{i,n}^{G/S} - P_{i,n-1}^{G/S} \leq rr_i \cdot T, \quad \forall i \in \{G,S\}, n, \quad (17)$$

$$-f_l \leq \sum_b SF_{l,b} \cdot \left( \boldsymbol{B}_b^G \cdot \boldsymbol{P}_n^G + \boldsymbol{B}_b^S \cdot \boldsymbol{P}_n^S + \boldsymbol{B}_b^{RE} \cdot \boldsymbol{P}_n^{RE} - P_{b,n}^{LD} \right) \leq f_l, \forall n,l, \quad (18)$$

where $\boldsymbol{P}_n^G$, $\boldsymbol{P}_n^S$, and $\boldsymbol{P}_n^{RE}$ are vectors composing $P_{i,n}^G$, $P_{i,n}^S$, and $P_{i,n}^{RE}$, respectively.

### 2) AGC constraints

Considering that the assessment of frequency regulation performance is based on the overall response results of the AGC signals, the AGC constraints below are formulated for an RTED time interval rather than an individual AGC period. Furthermore, the statistical variables of AGC signals are utilized to build the constraints.

The power output of the FRR after responding to the AGC signal should be within the operational limits, formulated as:

$$\mathbb{P}\left(P_{i,n}^{G/S} + PF_{i,n} \cdot MA_n^+ \leq P_{i,\max}\right) \geq \beta, \quad \forall i \in \{G,S\}, n, \quad (19\text{-a})$$

$$\mathbb{P}\left(P_{i,n}^{G/S} - PF_{i,n} \cdot MA_n^- \geq P_{i,\min}\right) \geq \beta, \quad \forall i \in \{G,S\}, n. \quad (19\text{-b})$$

The ramp rate of the FRR should be higher than the regulation rates of AGC, as:

$$\mathbb{P}\left(PF_{i,n} \cdot RR_n^+ \leq rr_i\right) \geq \beta, \quad \forall i \in \{G,S\}, n, \quad (20\text{-a})$$

$$\mathbb{P}\left(PF_{i,n} \cdot RR_n^- \leq rr_i\right) \geq \beta, \quad \forall i \in \{G,S\}, n. \quad (20\text{-b})$$

The FRR also needs to ramp between adjacent RTED operation points, which may limit the ramp rate for frequency regulation. Thus, the constraints below should be added, given as follows:

$$\mathbb{P}\left(PF_{i,n} \cdot RR_n^+ \leq rr_i - \frac{\left(P_{i,n}^{G/S} - P_{i,n-1}^{G/S}\right)}{T}\right) \geq \beta, \forall i \in \{G,S\}, n, \quad (21\text{-a})$$

$$\mathbb{P}\left(PF_{i,n} \cdot RR_n^- \leq rr_i - \frac{\left(P_{i,n}^{G/S} - P_{i,n-1}^{G/S}\right)}{T}\right) \geq \beta, \forall i \in \{G,S\}, n. \quad (21\text{-b})$$

The sum of the participation factors of all FRRs should equal 1, and each participation factor should be between 0 and 1, as follows:

$$\sum_{i \in \{S,G\}} PF_{i,n} = 1, \quad \forall n, \quad (22)$$

$$0 < PF_{i,n} < 1, \quad \forall i, n \quad (23)$$

### 3) SOC constraints of the ESS

In this research, the energy consumed by frequency regulation is considered when calculating the SOC change of the ESS. First, let $\kappa_n^+$ denote the proportion of the upward regulation signal in the $n^{th}$ RTED time interval:

$$\kappa_n^+ = \frac{X_n^+}{X_n^+ + X_n^-}. \quad (24)$$

$\kappa_n^+$ is also a stochastic variable related to AGC signals.

Secondly, the average power output of the ESS considering the energy spent on upward/downward frequency regulation can be calculated by using the base power of the ESS to add/subtract the average upward/downward regulation amplitude:

$$P_{i,n}^{S+} = P_{i,n}^S + \frac{PF_{i,n} \cdot E_n^+}{\kappa_n^+ \cdot T}, \quad (25\text{-a})$$

$$P_{i,n}^{S-} = P_{i,n}^S - \frac{PF_{i,n} \cdot E_n^-}{(1-\kappa_n^+) \cdot T}. \quad (25\text{-b})$$

Then, the SOC change of the $i^{th}$ battery during the $n^{th}$ RTED time interval can be calculated. There are three cases when calculating the SOC change.

1) The average output powers considering the energies spent on upward and downward frequency regulations are both larger than 0 (discharging), i.e., $P_{i,n}^{S+} \geq P_{i,n}^{S-} \geq 0$. Then, the SOC change can be calculated as

$$\Delta SOC_{i,n} = \frac{\kappa_n^+ \cdot P_{i,n}^{S+} \cdot T}{\eta_d \cdot CE_i} + \frac{(1-\kappa_n^+) \cdot P_{i,n}^{S-} \cdot T}{\eta_d \cdot CE_i} \triangleq L_1(n) \quad (26)$$

2) The average output powers considering the energies spent on upward and downward frequency regulations are respectively larger than 0 (discharging) and smaller than 0 (charging), respectively, i.e., $P_{i,n}^{S+} \geq 0 \geq P_{i,n}^{S-}$. Then, the SOC change can be calculated as

$$\Delta SOC_{i,n} = \frac{\kappa_n^+ \cdot P_{i,n}^{S+} \cdot T}{\eta_d \cdot CE_i} + \frac{(1-\kappa_n^+) \cdot P_{i,n}^{S-} \cdot T \cdot \eta_c}{CE_i} \triangleq L_2(n) \quad (27)$$

3) The average output powers considering the energies spent on upward and downward frequency regulations are both smaller than 0 (charging), i.e., $0 \geq P_{i,n}^{S+} \geq P_{i,n}^{S-}$. Then, the SOC change can be calculated as

$$\Delta SOC_{i,n} = \frac{\kappa_n^+ \cdot P_{i,n}^{S+} \cdot T \cdot \eta_c}{CE_i} + \frac{(1-\kappa_n^+) \cdot P_{i,n}^{S-} \cdot T \cdot \eta_c}{CE_i} \triangleq L_3(n) \quad (28)$$

**Proposition 1:** The SOC change (26)~(28) is equivalent to the equation below:

$$\Delta SOC_{i,n} = \max(L_1(n), L_2(n), L_3(n)). \quad (29)$$

The proof for Proposition 1 is given in Appendix B.

The SOC constraints at the end of the $n^{th}$ RTED time interval are:

$$\mathbb{P}\left(SOC_{i,n-1} - \Delta SOC_{i,n} \leq SOC_i^{\max}\right) \geq \beta, \quad \forall i \in S, n, \quad (30)$$

$$\mathbb{P}\left(SOC_{i,n-1} - \Delta SOC_{i,n} \geq SOC_i^{\min}\right) \geq \beta, \quad \forall i \in S, n. \quad (31)$$

As shown in (29), $\Delta SOC_{i,n}$ is the pointwise maximum of three linear functions; thus, constraints (30) and (31) should be satisfied for every segment of $\Delta SOC_{i,n}$, i.e., $L_1(n)$, $L_2(n)$, and $L_3(n)$. This will lead to an exponential explosion as $n$ increases, e.g., $n$ equal to 6 means there are $2 \times 3^6 = 1458$ SOC constraints. In addition, a new stochastic variable $\kappa_n^+$ is introduced in (30) and (31). Therefore, the original SOC constraints are conservatively reformulated as follows.

**Proposition 2:** The chance constraints (30)-(31) hold if the following constraints hold.

$$\mathbb{P}\left(SOC_{i,n-1} - \frac{P_{i,n}^S \cdot T + PF_{i,n} \cdot (E_n^+ - E_n^-)}{CE_i} \cdot \eta_c \leq SOC_i^{\max}\right) \geq \beta, \quad \forall i \in S, n, \quad (32)$$

$$\mathbb{P}\left\{SOC_{i,n-1} - \left(\frac{P_{i,n}^d \cdot T + PF_{i,n} \cdot E_n^+}{\eta_d \cdot CE_i} - \frac{P_{i,n}^c \cdot T + PF_{i,n} \cdot E_n^-}{CE_i} \cdot \eta_c\right) \geq SOC_i^{\min}\right\} \geq \beta,$$

$$\forall i \in S, n. \quad (33)$$

The proof for Proposition 2 is given in Appendix C.

## IV. DATA-DRIVEN DRO MODEL

Considering that the true distribution of $\xi^{\text{AGC}}$ cannot be forecasted precisely, the stochastic optimization model proposed in Section III is first reformulated as a DRO model in this section. Then, the Wasserstein metric-based DRO technique [13] is utilized to convert it to a linear programming model.

### A. Ambiguity Set of DRO

The objective of DRO is to find a decision that minimizes the worst-case expected cost. The worst case is taken over an ambiguity set constructed from the samples of $\xi^{\text{AGC}}$. In this paper, the ambiguity set is formulated by leveraging the Wasserstein metric. Compared with other approaches, Wasserstein balls provide an upper confidence bound, quantified by the Wasserstein radius $\varepsilon$, to achieve the superior out-of-sample performance [13]. Additionally, the Wasserstein metric-based DRO problem can be reformulated as a finite-dimensional linear programming problem, which can be solved by mature mathematical programming methods.

The uncertainty set is assumed to be a polytope $\Xi \triangleq \{\xi \in \mathbb{R}^{N_\xi} : C\xi \leq d\}$, where $N_\xi$ is the number of stochastic variables. The Wasserstein metric is defined on the space $\mathcal{S}(\Xi)$ composed of all probability distributions $\mathbb{Q}$ supported on $\Xi$ with

$$\mathbb{E}^{\mathbb{Q}}\|\xi\| = \int_{\Xi}\|\xi\|\mathbb{Q}(\text{d}\xi) < \infty. \quad (34)$$

**Definition** (Wasserstein Metric [34]): The Wasserstein metric $d_W : \mathcal{S}(\Xi) \times \mathcal{S}(\Xi) \to \mathbb{R}_+$ is defined for any distributions $\mathbb{Q}_1, \mathbb{Q}_2 \in \mathcal{S}(\Xi)$ via

$$d_W(\mathbb{Q}_1,\mathbb{Q}_2) = \sup_{f \in \mathbb{L}}\left\{\int_{\Xi} f(\xi)\mathbb{Q}_1(\text{d}\xi) - \int_{\Xi} f(\xi)\mathbb{Q}_2(\text{d}\xi)\right\}, \quad (35)$$

where $\mathbb{L}$ denotes the spaces of all Lipschitz functions with $|f(\xi) - f(\xi')| \leq \|\xi - \xi'\|$ for all $\xi, \xi' \in \Xi$.

We can now define the ambiguity set of the distribution of $\xi^{\text{AGC}}$ as:

$$\widehat{\mathcal{P}} \triangleq \left\{\mathbb{Q} \in \mathcal{S}(\Xi) : d_w(\mathbb{Q},\widehat{\mathbb{P}}) \leq \varepsilon\right\}, \quad (36)$$

where $\mathbb{Q}$ denotes a distribution of $\xi^{\text{AGC}}$ defined on space $\mathcal{S}(\Xi)$, $\widehat{\mathbb{P}}$ is the empirical distribution of $\xi^{\text{AGC}}$ estimated based on the samples taken from the Copula-based joint distribution model, and $\varepsilon$ is the radius of the Wassenstein ball. The ambiguity set $\widehat{\mathcal{P}}$ can be viewed as a Wassenstein ball of radius $\varepsilon$ centered at the empirical distribution $\widehat{\mathbb{P}}$. By properly choosing $\varepsilon$, the true distribution $\mathbb{P}$ of $\xi^{\text{AGC}}$ can be included in $\widehat{\mathcal{P}}$ with a prescribed confidence level, and the robustness of the optimization result can be guaranteed [34].

### B. Reformulation of the optimization model

The objective of DRO in this study is to find the generation plan $[P^G, P^d, P^c]$ for RTED and the participation factors $PF$ for frequency regulation that minimize the total cost of power generation and frequency regulation under the most adverse distribution of $\xi^{\text{AGC}}$ within the ambiguity set $\widehat{\mathcal{P}}$. By using the DRO theory, the optimization model proposed in Section III can be reformulated as the following compact form (denoted as $\mathcal{P}1$). For ease of notation, most of the subscripts $n$, which denotes the index of the RTED time interval, are omitted, except for those in constraints (21), (32), and (33).

$$\min\left\{\mathcal{A}(P^c, P^d) + \max_{\mathbb{Q}\in\widehat{\mathcal{P}}}\mathbb{E}^{\mathbb{Q}}\sum_{i\in G}\max_{k_i\leq K}\left[\mathcal{B}_{k_i}(PF_i)\cdot\xi^{\text{AGC}\top}\right.\right.$$
$$\left.\left.+\mathcal{D}_{k_i}(P_i^G)\right] + \max_{\mathbb{Q}\in\widehat{\mathcal{P}}}\mathbb{E}^{\mathbb{Q}}\mathcal{L}(PF)\cdot\xi^{\text{AGC}\top}\right\},$$
$$\text{s.t.} \quad \min_{\mathbb{Q}\in\widehat{\mathcal{P}}}\mathbb{P}\left[\mathcal{M}_j(PF_i)\cdot\xi^{\text{AGC}\top}+\mathcal{N}_j(P_i^{G/S},P_i^c,P_i^d)\leq 0\right]\geq\beta_j,$$
$$\forall i \in \{G, S\}, j \in \mathcal{J}$$
constraints (15)~(18),(22),(23)
$$(37)$$

where $\mathcal{A}(P^c, P^d)$ denotes the total degradation cost of all ESSs, $\max_{\mathbb{Q}\in\widehat{\mathcal{P}}}\mathbb{E}^{\mathbb{Q}}\sum_{i\in G}\max_{k_i\leq K}[\mathcal{B}_{k_i}(PF_i)\cdot\xi^{\text{AGC}\top}+\mathcal{D}_{k_i}(P_i^G)]$ is the total generation cost of all generators, and $\max_{\mathbb{Q}\in\widehat{\mathcal{P}}}\mathbb{E}^{\mathbb{Q}}\mathcal{L}(PF)\cdot\xi^{\text{AGC}\top}$ is the total regulation cost of all FRRs. $j$ and $\mathcal{J}$ are the index and set of chance constraints, respectively. We slightly abuse the index symbol $j$, but readers can easily distinguish its implications. Other uppercase calligraphic letters in (37) are functions, and their explicit definitions are given in Appendix D.

The chance constraints in $\mathcal{P}1$ are reformulated by leveraging the CVaR approximation [35] in this research. CVaR is a risk measure that quantifies the expected loss over the part of distribution beyond the confidence level, which has been widely used in risk management and stochastic optimization [36]. After transformation, the chance constraint can be transformed into:

$$\max_{\mathbb{Q}\in\widehat{\mathcal{P}}} \text{CVaR}_\beta\left[\mathcal{M}_j(PF_i)\cdot\xi^{\text{AGC}\top}+\mathcal{N}_j(P_i^{G/S},P_i^c,P_i^d)\right]$$
$$= \max_{\mathbb{Q}\in\widehat{\mathcal{P}}}\left[\delta_j^0+\frac{1}{1-\beta}\cdot\mathbb{E}^{\mathbb{Q}}\left(\mathcal{M}_j(PF_i)\cdot\xi^{\text{AGC}\top}+\mathcal{N}_j(P_i^{G/S},P_i^c,P_i^d)-\delta_j^0\right)^+\right]$$
$$\triangleq \max_{\mathbb{Q}\in\widehat{\mathcal{P}}}\mathbb{E}^{\mathbb{Q}}\max_{k=1,2}\left[\mathcal{M}'_{j,k}(PF_i)\cdot\xi^{\text{AGC}\top}+\mathcal{N}'_{j,k}(P_i^{G/S},P_i^c,P_i^d,\delta_j^0)\right],$$
$$(38)$$

where $\delta_j^0$ denotes the value at risk of the $j^{th}$ chance constraint. In this work, the CVaR-reformulated chance constraints are multiplied by a risk aversion parameter $\rho$ and added into the objective function. Then, $\mathcal{P}1$ can be reformulated as follows (denoted as $\mathcal{P}2$), which can also be viewed as a risk-aversion optimization problem.

$$\min\left\{\mathcal{A}(P^c,P^d)+\max_{\mathbb{Q}\in\widehat{\mathcal{P}}}\mathbb{E}^{\mathbb{Q}}\sum_{i\in G}\max_{k_i\leq K}\left[\mathcal{B}_{k_i}(PF_i)\cdot\xi^{\text{AGC}\top}+\mathcal{D}_{k_i}(P_i^G)\right]\right.$$
$$+\max_{\mathbb{Q}\in\widehat{\mathcal{P}}}\mathbb{E}^{\mathbb{Q}}\mathcal{L}(PF)\cdot\xi^{\text{AGC}\top}+\rho\cdot\sum_{j\in\mathcal{J}}\sum_{i\in\{G,S\}}\max_{\mathbb{Q}\in\widehat{\mathcal{P}}}\mathbb{E}^{\mathbb{Q}}\max_{k=1,2}\left[\mathcal{M}'_{j,k}(PF_i)\cdot\xi^{\text{AGC}\top}\right.$$
$$\left.\left.+\mathcal{N}'_{j,k}(P_i^{G/S},P_i^c,P_i^d,\delta_j^0)\right]\right\},$$
s.t. constraints (15)~(18),(22),(23)
$$(39)$$

*Remark 2*: Instead of adding the CVaR-reformulated chance constraints to the objective function, transforming them as hard constraints is also technically feasible [18], [37]. We consider that the treatment of CVaR-approximation depends on the real engineering meanings behind it. The chance constraints in this study concern the secondary frequency control, which mainly influences the frequency control performance rather than the stability of the power system. Consequently, we consider that there is no need to pursue a high frequency control performance all the time if it comes with a very high regulation cost. Adding the CVaR-approximated chance constraints can balance the frequency control performance and regulation cost, which is better for the social benefit. It also provides more flexibility for power system operators under different risk preference, which can be achieved by choosing different risk-aversion parameter (will be illustrated in Section V.E). Thus, the CVaR approximations are added into the objective function in this work. Nevertheless, if the violations of these chance constraints are not permitted for the control area, the CVaR-reformulated chance constraints should be enforced as hard constraints. The interested redears are refered to [18] for the detailed transformation, where two tractable formulations of CVaR-approximated chance constraint are rigorously developed.

The difficulty of solving $\mathcal{P}2$ lies in the expectation terms. According to [13], the DRO with the worst-case expectation of the piecewise linear affine function can be transformed into a computationally tractable linear programming model. However, the total generation cost term

$$\max_{\mathbb{Q}\in\widehat{\mathcal{P}}}\mathbb{E}^{\mathbb{Q}}\sum_{i\in G}\max_{k_i\leq K}\left[\mathcal{B}_{k_i}(PF_i)\cdot\boldsymbol{\xi}^{\mathrm{AGC}\top}+\mathcal{D}_{k_i}(P_i^{\mathrm{G}})\right]$$

is a summation over numbers of piecewise linear affine functions. Theoretically, it can be reformulated as one piecewise linear affine function [13], and the summation of the $N$ piecewise functions should be formulated as the maximum of $K^N$ functions. In this research, $N$ and $K$ are equal to 19 and 3 respectively, and the new piecewise function is the maximum of $3^{19}$ functions, indicating that an efficient solution may not be available. To solve this problem, we exchange the summation $\sum_{i\in G}$ and the operator $\max_{\mathbb{Q}\in\widehat{\mathcal{P}}}\mathbb{E}^{\mathbb{Q}}$ in $\mathcal{P}2$ as

$$\sum_{i\in G}\max_{\mathbb{Q}\in\widehat{\mathcal{P}}}\mathbb{E}^{\mathbb{Q}}\max_{k_i\leq K}\left[\mathcal{B}_{k_i}(PF_i)\cdot\boldsymbol{\xi}^{\mathrm{AGC}\top}+\mathcal{D}_{k_i}(P_i^{\mathrm{G}})\right]$$

This reformulation is a conservative approximation of the original total generation cost, which may influence the optimization results. Thus, it is tested and analyzed in Section V.C.

After this transformation, all the expectation terms in the objective function share the same piecewise linear affine form. By using $\mathcal{U}$ to denote $\{\mathcal{B}(PF_i), \mathcal{L}(PF), \mathcal{M}'(PF_i)\}$, $\mathcal{V}$ to denote $\{\mathcal{D}(P_i^{\mathrm{G}}), 0, \mathcal{N}'(P_i^{\mathrm{G/S}},P_i^{\mathrm{c}},P_i^{\mathrm{d}},\delta_j^0)\}$, and $h$ and $\mathcal{H}$ to denote the index and set of the expectation terms in $\mathcal{P}2$, respectively, the objective function in $\mathcal{P}2$ can be expressed as

$$\min\left\{\mathcal{A}(\boldsymbol{P}^{\mathrm{c}},\boldsymbol{P}^{\mathrm{d}})+\max_{\mathbb{Q}\in\widehat{\mathcal{P}}}\mathbb{E}^{\mathbb{Q}}\sum_{h\in\mathcal{H}}\max_{k_h\leq K_h}\left(\mathcal{U}_{k_h}\cdot\boldsymbol{\xi}^{\mathrm{AGC}\top}+\mathcal{V}_{k_h}\right)\right\},$$

where $k_h$ and $K_h$ are the index and the total number of segments of the $h^{\mathrm{th}}$ piecewise linear function, respectively. Then, $\mathcal{P}2$ can be transformed as the linear programming problem below (denoted as $\mathcal{P}3$).

$$\begin{aligned}
\min\quad & \mathcal{A}(\boldsymbol{P}^{\mathrm{c}},\boldsymbol{P}^{\mathrm{d}})+\sum_{h\in\mathcal{H}}\left(\lambda_h\varepsilon+\frac{1}{\Upsilon}\sum_{\upsilon=1}^{\Upsilon}s_{h,\upsilon}\right),\\
\text{s.t.}\quad & \mathcal{U}_{k_h}\cdot\widehat{\boldsymbol{\xi}_\upsilon^{\mathrm{AGC}}}^\top+\mathcal{V}_{k_h}+\gamma_{k_h,\upsilon}\cdot\left(\boldsymbol{d}-\boldsymbol{C}\cdot\widehat{\boldsymbol{\xi}_\upsilon^{\mathrm{AGC}}}^\top\right)\leq s_{h,\upsilon},\forall h,k,\upsilon,\\
& \left\|\boldsymbol{C}^\top\gamma_{k_h,\upsilon}-\mathcal{U}_{k_h}\right\|_\infty\leq\lambda_h,\qquad\forall h,k,\upsilon,\\
& \gamma_{k_h,\upsilon}\geq 0,\qquad\forall h,k,\upsilon,\\
& \text{constraints (15)}\sim\text{(18),(22),(23)}
\end{aligned}$$

(40)

where $\lambda_h$, $s_{h,\upsilon}$ and $\gamma_{k_h,\upsilon}$ are auxiliary variables, $\upsilon$ and $\Upsilon$ are the index and the total number of samples, respectively, $\widehat{\boldsymbol{\xi}^{\mathrm{AGC}}}$ represents the sample of $\boldsymbol{\xi}^{\mathrm{AGC}}$, and $\|\cdot\|_\infty$ is the $\infty$-norm.

$\mathcal{P}3$ is a linear programming model that can be efficiently solved by mature mathematical programming methods.

## V. Case Study

### A. Case Settings

The proposed method requires historical data of the AGC

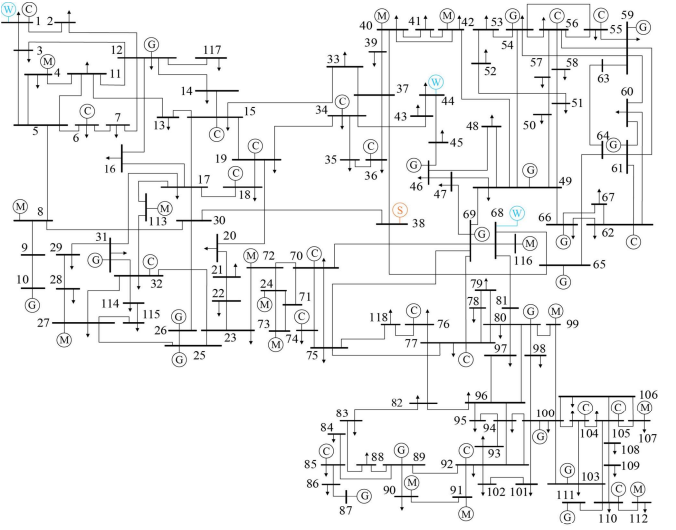

Fig. 3. Diagram of the IEEE 118-bus test system.

signal, load power, and renewable generations. The data from a real power system in North China are used for the case study, and all the historical data are scaled down for the simulations. The data from January 2016 to February 2018 are used to build the Copula-based AGC signal model. The data of March 2018 are utilized to test the proposed method.

The modified IEEE 118-bus system is used as the test system [38], as shown in Fig. 3. Three wind farms are connected to buses 1, 44, and 68, with installed capacities of 400 MW, 300 MW, and 300 MW, respectively; one photovoltaic power station with an installed capacity of 330 MW is connected to bus 38. The ramp rates and regulation mileage costs of the generators are listed in Table I, which are taken from [39] and [7], respectively. The parameters of the ESSs are shown in Table II.

TABLE I Generator Parameters Used in The Case Studies

| Generator (# bus) | Ramp rate (MW/s) | Mileage cost ($/MW) |
|---|---|---|
| 10,12,25,26 | 0.05 | 2 |
| 31,46,49,54 | 0.1 | 3 |
| 59,61,65,66 | 0.2 | 4 |
| 69,80,87,89 | 0.25 | 5 |
| 100,103,111 | 0.4 | 10 |

TABLE II ESS Parameters Used In The Case Studies

| Connected bus | 4 | 9 | 39 | 67 |
|---|---|---|---|---|
| Power capacity (MW) | 10 | 20 | 10 | 20 |
| Energy capacity (MWh) | 5 | 10 | 5 | 10 |
| Ramp rate (MW/s) | 4 | 4 | 4 | 4 |
| Degradation cost ($/MWh) | 200 | 200 | 200 | 200 |
| Regulation mileage cost ($/MW) | 10 | 10 | 10 | 10 |
| Initial SOC | 10% | 50% | 90% | 50% |
| Maximum SOC | 90% | 90% | 90% | 90% |
| Minimum SOC | 10% | 10% | 10% | 10% |

### B. Copula Model Selection Result

In this research, Gaussian Copula and Student-t Copula in the Elliptical Copula family, as well as Clayton Copula, Gumbel Copula, and Frank Copula in the Archimedean Copula family are applied to construct the Copula-based AGC signal model [40]. The model performances are evaluated with the Bayesian information criterion (BIC), which is a criterion for model selection among a finite set of models. The BIC is formally defined as [41]:

$$BIC=-2\ln(\ell)+q*\ln(\mathcal{X}),\qquad(41)$$

where $q$ denotes the number of parameters estimated by the Copula model, which is different for each kind of Copula function, $\ell$ indicates the value of the maximum likelihood function [42], and $\mathcal{N}$ is the total number of samples used in the model development. The first part of the BIC measures the likelihood of a model: a larger $\ell$ contributes to a smaller BIC. The second part aims to avoid the overfitting problem by introducing a penalty term for the number of parameters in the model: a smaller $q$ results in a lower BIC. Thus, the model with the lowest BIC is preferred.

The performances of each model are shown in Table III. It can be observed that the Student-t Copula model has the lowest BIC, so it is applied to build the AGC signal model.

TABLE III PERFORMANCES OF DIFFERENT COPULA MODELS

| Model | Gaussian | Student-t | Gumbel | Clayton | Frank |
|---|---|---|---|---|---|
| BIC ($10^3$) | -1825.22 | -1991.53 | -3.29 | -23.30 | -4.22 |

To test the performance of the proposed model, the conditional PDFs estimated by the real distribution and proposed model are compared. First, the power variation ranges $[\underline{\Delta P^{LD}}, \overline{\Delta P^{LD}}]$, $[\underline{\Delta P^{PV}}, \overline{\Delta P^{PV}}]$, and $[\underline{\Delta P^{WP}}, \overline{\Delta P^{WP}}]$ are selected as the condition of the distribution of AGC signals. Then, some data samples in the test set (in March 2018) are selected with their power variations belonging to the given ranges, i.e., $\Delta P_\theta^{LD} \in [\underline{\Delta P^{LD}}, \overline{\Delta P^{LD}}]$, $\Delta P_\theta^{PV} \in [\underline{\Delta P^{PV}}, \overline{\Delta P^{PV}}]$, and $\Delta P_\theta^{WP} \in [\underline{\Delta P^{WP}}, \overline{\Delta P^{WP}}]$ for every selected data sample denoted by $\theta$. Thirdly, the Copula model is utilized to generate samples conditioned on the power variations within the selected ranges. Finally, the normal kernel function is used to build the probability densities of the selected historical data samples and the generated data samples. In the first step, the width of each interval is fixed, and the group of power variation intervals including the most data samples in the test set is used for analysis. The conditional PDFs for the quantile of upward regulation amplitude are given as an example, and the comparison results are shown in Fig. 4.

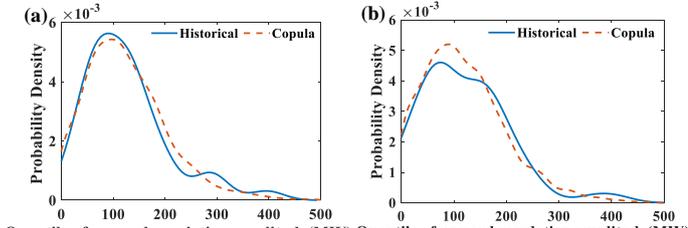
Fig. 4. Conditional PDFs for the quantile of upward regulation amplitude: (a) for the daytime, (b) for the night.

The left figure shows the daytime situation, where the ranges of the load power, wind generation, and solar power variations are [16.72, 201], [-27.22, 22.83], and [-41.11, -34.14], respectively. The right figure illustrates the situation at night, where the solar power variation is 0. The ranges of the load power and wind generation variations are [-32.26, 132] and [-17.2, 28.35], respectively. The results show that, under both circumstances, the proposed model provides reasonable conditional PDFs of the statistical variables of AGC signals that are close to the real statistical distributions.

### C. Comparison of Different Methods

In this section, the proposed method is compared with the existing robust and traditional methods:

*1) Proposed method*

After calculating $\Delta \boldsymbol{P}_n$ based on the forecasted load power and renewable generations, 30 samples are taken according to (9) for each RTED time interval in the optimization horizon. These samples are used in the proposed DRO model. The radius of the Wasserstein ball is selected as 0.3 through cross-validation [13]. The $\boldsymbol{C}$ and $\boldsymbol{d}$ in $\mathcal{P}3$ are set as $\boldsymbol{0}$ in this research, as in [43], [44], which assumes that there are no bounds on the statistical variables of the AGC signals.

*2) Robust optimization method*

The statistical variables of the AGC signals are forecasted by using the extreme learning machine [45]. The SOC constraints of the ESS adopt the constraints proposed in [10]. The forecasted quantiles of regulation amplitudes and regulation rates are substituted into other AGC constraints, i.e., constraints (19) through (21), to formulate the hard constraints.

*3) Traditional method*

The secondary frequency regulation cost and constraint are not considered in the RTED optimization. The participation factor of each FRR is calculated according to the proportion of its regulation capacity to the total regulation capacity.

We use the aforementioned methods to sequentially optimize the generation plan $[\boldsymbol{P}^G, \boldsymbol{P}^d, \boldsymbol{P}^c]$ for RTED and the participation factors $\boldsymbol{PF}$ for frequency regulation in the test timespan. The RTED timespan is five minutes in this study, and the horizon for the rolling optimization model is half an hour, i.e., six RTED time intervals. To test and compare the optimization results of each method, the historical data in the test timespan are used to calculate $\boldsymbol{\xi}^{AGC}$ according to Subsection II. Then, $\boldsymbol{\xi}^{AGC}$ are substituted into the objective functions in (10) to calculate the operational costs and taken into the chance constraints to calculate the penalty on constraint violation. The total cost is the sum of the operational cost and the penalty on constraint violation. The risk aversion parameter $\rho$ is set as 15 for all methods.

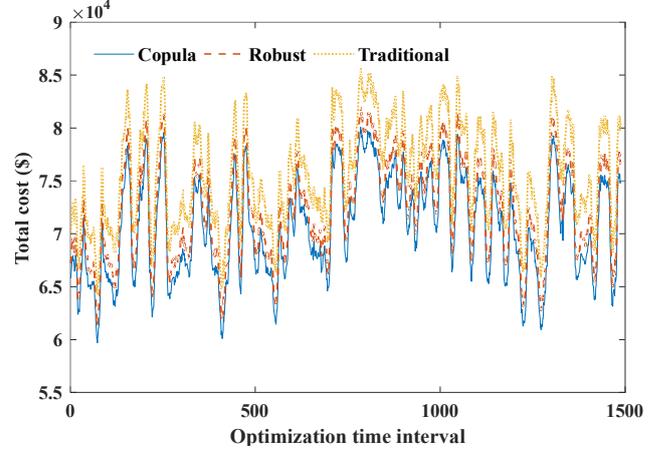
Fig. 5. Comparison of the total costs obtained by different methods.

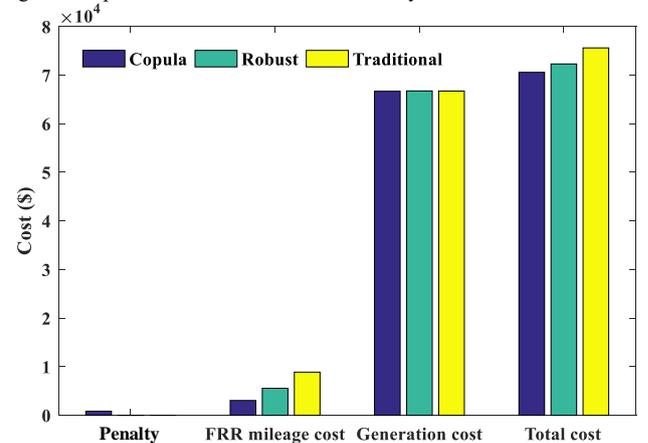
Fig. 6. Cost components obtained by different optimization methods.

The total costs of these methods are illustrated in Fig. 5. The cost of the proposed method is less than or equal to those of other methods in the whole test timespan, which proves the superiority of the proposed method.

The cost components obtained by different optimization methods are shown in Fig. 6. The frequency regulation cost and the total cost of the traditional method are higher than those of the other methods. Moreover, the generation cost of the traditional method is slightly higher than that of the proposed method, as shown in Table IV, because the traditional method optimizes the generation plan without considering the generation cost consumed by the frequency regulation. Therefore, it is necessary to consider the secondary frequency regulation in RTED.

In addition, the comparison of the generation costs proves that the conservative approximation of the total generation cost made in the DRO reformulation is acceptable in this study. Because even after the conservative approximation, the generation cost of the proposed method is still the smallest.

TABLE IV COMPARISON OF THE GENERATION COSTS OF DIFFERENT METHODS

| Method | Proposed | Robust | Traditional |
|---|---|---|---|
| Generation cost ($) | 66687 | 66753 | 66724 |

The penalty on constraint violation of the robust method is lower than that of the proposed method, but the FRR mileage cost and generation cost of the former are higher than those of the proposed method. Furthermore, the total cost of the proposed method is lower than that of the robust optimization method. These comparison results show that the conservativeness of the proposed data-driven DRO method is lower than that of the robust optimization method.

### D. Comparison of Computational Efficiency

The experiments are performed on a PC with an Intel(R) Core(TM) i7-7700 CPU 3.6 GHz and 8 GB of memory. The algorithms are implemented in MATLAB and programmed using YALMIP [46]. The linear programming solver is MOSEK 9.1.13 [47]. The computation times for the different methods are shown in Table V. The DRO method has the longest computation time. However, RTED is typically executed every five minutes; thus, the computation time of the proposed method is acceptable.

TABLE V COMPUTATION TIME FOR DIFFERENT METHODS

| Method | DRO | Robust | Traditional |
|---|---|---|---|
| Computation time (s) | 1.46 | 0.82 | 0.45 |

### E. Sensitivity Analysis

*1) Risk aversion parameter $\rho$*

In this research, the risk aversion preference of the system operator can be controlled by adjusting $\rho$. To perform sensitivity analysis, the parameter $\rho$ is changed from 15 to 75 with a step size of 30. Fig. 7 illustrates the optimization results of the proposed method with different $\rho$.

As $\rho$ increases, the penalty on constraint violation decreases, while the frequency regulation cost and the total cost increase. The penalty on constraint violation is very small when $\rho$ equals 75. The results prove that the trade-off between the frequency control performance and regulation cost can be controlled by adjusting $\rho$.

*2) Sample size for DRO*

To analyze the influence of the sample size on the DRO result, the sample size is increased from 30 to 120 with a step size of 30. The total cost and solving time under each sample size are listed in Table VI.

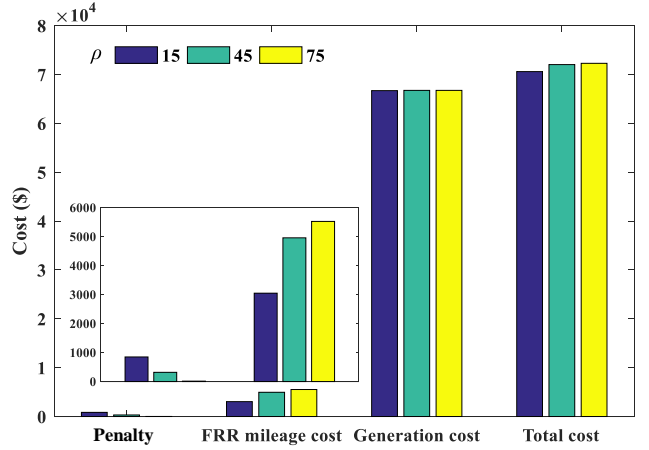

Fig. 7. Optimization results of the proposed method under different values of the parameter $\rho$.

TABLE VI TOTAL COST AND COMPUTATION TIME UNDER DIFFERENT SAMPLE SIZES

| Sample size | 30 | 60 | 90 | 120 |
|---|---|---|---|---|
| Total cost ($10^4$ $) | 7.0362 | 7.0349 | 7.0345 | 7.0330 |
| Computation time (s) | 1.46 | 2.36 | 3.12 | 4.76 |

It can be seen that the total cost decreases with increasing sample size, while the computation time increases during this process. The total cost decreases because the ambiguity set shrinks and the worst-case distribution in the ambiguity set becomes closer to the true distribution as more samples are incorporated. The calculation time increases because the number of variables and constraints in the optimization problem increase as the sample size increases.

## VI. CONCLUSION

This paper proposed a data-driven DRO model for the RTED considering the frequency regulation cost and constraint. The correlation among the AGC signal, load power and renewable generation variations was considered via the Copula theory-based AGC signal model. The optimization model was transformed into a linear programming problem by leveraging the Wasserstein metric-based DRO technique. The simulation results show that the following:
1) Considering the frequency regulation cost is necessary for RTED, which can decrease the total operational cost compared with the traditional method which does not consider it.
2) The conservativeness of the proposed data-driven DRO method is lower than that of the robust optimization method, contributing to a lower total operational cost.

In future work, studies will be carried out to further investigate the coordination of FRRs under different response times.

## APPENDIX A

Multivariate Gaussian Copula is a widely used Copula function. Let $\varpi$ be a symmetric, positive definite matrix with $\text{diag}(\varpi) = \mathbf{1}$, $\Psi_\varpi$ be the standardized multivariate normal distribution with correlation matrix $\varpi$, and $\Psi^{-1}$ be the inverse function of the normal distribution. The multivariate Gaussian Copula for the stochastic variables $\{u_1,...,u_z,...,u_Z\}$ that follow the uniform distribution is then defined as follows[12]:

$$\mathcal{C}(u_1,...,u_z,...,u_Z;\varpi) = \Psi_\varpi\left(\Psi^{-1}(u_1),...,\Psi^{-1}(u_z),...,\Psi^{-1}(u_Z)\right)$$

And the corresponding density function is:

$$\varsigma(u_1,\ldots,u_z,\ldots,u_Z;\varpi) = \frac{1}{|\varpi|^{1/2}} \exp\left(-\frac{1}{2}\psi^\top(\varpi^{-1}-I_Z)\psi\right) \quad (43)$$

where $\psi$ is a vector composed of $\psi_z$, with $\psi_z = \Psi^{-1}(u_z)$, and $I_Z$ is the unit matrix with dimension of $Z$.

Other Copula functions can also be found in [12].

## APPENDIX B

To prove Proposition 1, we need to prove that $\max(L_1(n), L_2(n), L_3(n))$ will take the respective value under the three cases listed by (26)~(28).

In the case of $P_{i,n}^{S+} \geq P_{i,n}^{S-} \geq 0$, we have

$$\frac{\kappa_n^+ \cdot P_{i,n}^{S+} \cdot T}{\eta_d \cdot CE_i} \geq \frac{\kappa_n^+ \cdot P_{i,n}^{S+} \cdot T \cdot \eta_c}{CE_i} \quad (44)$$

and

$$\frac{(1-\kappa_n^+) \cdot P_{i,n}^{S-} \cdot T}{\eta_d \cdot CE_i} \geq \frac{(1-\kappa_n^+) \cdot P_{i,n}^{S-} \cdot T \cdot \eta_c}{CE_i} \quad (45)$$

Thus, $L_1(n) \geq L_2(n)$ and $L_1(n) \geq L_3(n)$ under this case. The proofs for the other two cases are similar.

## APPENDIX C

First, we prove that constraint (32) is a safe approximation of the original constraint (30).

By substituting (29) into (30), we have:

$$SOC_{i,n-1} - \Delta SOC_{i,n-1} = SOC_{i,n-1} - \max\{L_1(n), L_2(n), L_3(n)\} \leq SOC_{i,n-1} - \max\{L_1(n), L_3(n)\}. \quad (46)$$

When $0 \geq P_{i,n}^{S+} \geq P_{i,n}^{S-}$, equation (46) can be written as follows:

$$SOC_{i,n-1} - \max\{L_1(n), L_3(n)\} = SOC_{i,n-1} - L_3(n). \quad (47)$$

When $P_{i,n}^{S+} \geq P_{i,n}^{S-} \geq 0$, both $L_1(n)$ and $L_3(n)$ are larger than 0, which means that the ESS is discharging during the $n^{th}$ RTED time interval. Then, we can deduce easily from

$$SOC_{i,n-1} - L_3(n) \leq SOC_{i,\max} \quad (48)$$

to

$$SOC_{i,n-1} - \max\{L_1(n), L_3(n)\} \leq SOC_{i,\max}. \quad (49)$$

Based on the above analysis, the original maximum SOC constraints (30) can be conservatively reformulated to (32).

Secondly, we prove that the transformation from constraint (31) to (33) is also conservative.

By substituting (25) into $L_2$ in (27), we can reformulate $L_2$ as follows:

$$L_2 = \frac{\kappa_n^+ \cdot P_{i,n}^S \cdot T + PF_{i,n}^+ \cdot E_n^+}{\eta_d \cdot CE_i} + \frac{(1-\kappa_n^+) \cdot P_{i,n}^S \cdot T - PF_{i,n}^- \cdot E_n^-}{CE_i} \cdot \eta_c \quad (50)$$

When $P_{i,n}^S \geq 0$, we can reformulate (50) as

$$L_2 \leq \frac{\kappa_n^+ \cdot P_{i,n}^S \cdot T + PF_{i,n}^+ \cdot E_n^+}{\eta_d \cdot CE_i} + \frac{(1-\kappa_n^+) \cdot P_{i,n}^S \cdot T}{\eta_d \cdot CE_i} - \frac{PF_{i,n}^- \cdot E_n^-}{CE_i} \cdot \eta_c \quad (51)$$

$$= \frac{P_{i,n}^S \cdot T + PF_{i,n}^+ \cdot E_n^+}{\eta_d \cdot CE_i} - \frac{PF_{i,n}^- \cdot E_n^-}{CE_i} \cdot \eta_c,$$

which is denoted as $L_{2,1}$. When $P_{i,n}^S < 0$, we can reformulate (50) as follows:

$$L_2 < \frac{PF_{i,n}^+ \cdot E_n^+}{\eta_d \cdot CE_i} + \frac{\rho_n^+ \cdot P_{i,n}^S \cdot T}{CE_i} \cdot \eta_c + \frac{(1-\rho_n^+) \cdot P_{i,n}^S \cdot T - PF_{i,n}^- \cdot E_n^-}{CE_i} \cdot \eta_c \quad (52)$$

$$= \frac{PF_{i,n}^+ \cdot E_n^+}{\eta_d \cdot CE_i} + \frac{P_{i,n}^S \cdot T - PF_{i,n}^- \cdot E_n^-}{CE_i} \cdot \eta_c,$$

which is denoted as $L_{2,2}$. It can be deduced easily that:

$$L_2 \leq \max(L_{2,1}, L_{2,2}). \quad (53)$$

Comparing $L_{2,1}$ and $L_1$, $L_{2,2}$ and $L_3$ separately, it is easy to find that $L_{2,1} \geq L_1$, $L_{2,2} \geq L_3$. Thus, we have

$$\max\{L_1, L_2, L_3\} \leq \max(L_{2,1}, L_{2,2}) \quad (54)$$

Assuming that $P_{i,n}^d$ and $P_{i,n}^c$ cannot be positive at the same time, we have

$$\max(L_{2,1}, L_{2,2}) = \frac{P_{i,n}^d \cdot T + PF_{i,n}^+ \cdot E_n^+}{\eta_d \cdot CE_i} - \frac{P_{i,n}^c \cdot T + PF_{i,n}^- \cdot E_n^-}{CE_i} \cdot \eta_c \quad (55)$$

This assumption can be guaranteed by including the degradation cost in the objective function.

Thus, the original minimum SOC constraint (31) can be safely approximated by constraint (33). ∎

## APPENDIX D

To clarify the relationship between the compact form of the DRO model and the model developed in Section III, the definitions of the functions denoted by the uppercase calligraphic letters are given as follows.

$$\mathcal{A}(\boldsymbol{P}^c, \boldsymbol{P}^d) = \sum_{i \in S} \phi_i^S \cdot (P_i^d \eta_d^{-1} + P_i^c \eta_c) \cdot T$$

$$\mathcal{B}_{k_i}(PF_i) = [\phi_{i,k}^G \cdot PF_i, -\phi_{i,k}^G \cdot PF_i, 0, 0, 0, 0], \forall i \in G$$

$$\mathcal{D}_{k_i}(P_i^G) = (\phi_{i,k}^G \cdot P_i^G + \varphi_{i,k}^G) \cdot T, \quad \forall i \in G$$

$$\mathcal{L}(\boldsymbol{PF}) = \left[0, 0, \sum_{i \in \{G,S\}} \phi_i^R \cdot PF_i, 0, 0, 0, 0\right],$$

$$\mathcal{M}_j(PF_i) = \begin{cases} [0,0,0,PF_i,0,0,0] &, j=1 \text{ for (19.a)} \\ [0,0,0,0,PF_i,0,0] &, j=2 \text{ for (19.b)} \\ [0,0,0,0,0,PF_i,0] &, j=3 \text{ for (20.a)} \\ [0,0,0,0,0,0,PF_i] &, j=4 \text{ for (20.b)} \\ [0,0,0,0,0,PF_i,0] &, j=5 \text{ for (21.a)} \\ [0,0,0,0,0,PF_i,0] &, j=6 \text{ for (21.b)} \\ \left[-\dfrac{PF_i \cdot \eta_c}{CE_i}, \dfrac{PF_i \cdot \eta_c}{CE_i}, 0,0,0,0,0\right] &, j=7 \text{ for (32)} \\ \left[\dfrac{PF_i}{\eta_d \cdot CE_i}, -\dfrac{PF_i \cdot \eta_c}{CE_i}, 0,0,0,0,0\right] &, j=8 \text{ for (33)} \end{cases}$$

$$\mathcal{N}_j(P_i^{G/S}, P_i^c, P_i^d) = \begin{cases} P_{i,\max} - P_i^{G/S}, &, j=1 \text{ for (19.a)} \\ P_i^{G/S} - P_{i,\min}, &, j=2 \text{ for (19.b)} \\ rr_i, &, j=3 \text{ for (20.a)} \\ rr_i, &, j=4 \text{ for (20.b)} \\ rr_i - \dfrac{(P_{i,n}^{G/S} - P_{i,n-1}^{G/S})}{T}, &, j=5 \text{ for (21.a)} \\ rr_i - \dfrac{(P_{i,n}^{G/S} - P_{i,n-1}^{G/S})}{T}, &, j=6 \text{ for (21.b)} \\ SOC_i^{\max} - SOC_{i,n-1} + \dfrac{P_{i,n}^S \cdot T}{CE_i} \cdot \eta_c, &, j=7 \text{ for (32)} \\ SOC_{i,n-1} - \left(\dfrac{P_{i,n}^d \cdot T}{\eta_d \cdot CE_i} - \dfrac{P_{i,n}^c \cdot T \cdot \eta_c}{CE_i}\right) - SOC_i^{\min}, &, j=8 \text{ for (33)} \end{cases}$$

The definitions of the functions $\mathcal{M}'_{j,k}(PF_i)$ and

$\mathcal{N}'_{j,k}(P_i^{\text{G/S}}, P_i^{\text{c}}, P_i^{\text{d}}, \delta_j^0)$ in (38) are given as follow:

$$\mathcal{M}'_{j,k}(PF_i) = \begin{cases} \dfrac{\mathcal{M}_j(PF_i)}{1-\beta}, & k=1 \\ 0, & k=2 \end{cases}$$

$$\mathcal{N}'_{j,k}(P_i^{\text{G/S}}, P_i^{\text{c}}, P_i^{\text{d}}, \delta_j^0) = \begin{cases} \dfrac{\mathcal{N}_j(P_i^{\text{G/S}}, P_i^{\text{c}}, P_i^{\text{d}})}{1-\beta} - \dfrac{\beta \cdot \delta_j^0}{1-\beta}, & k=1 \\ \delta_j^0, & k=2 \end{cases}$$